\documentclass[11pt,a4paper]{article}
\usepackage{jheppub}
\usepackage{tikz}
\usetikzlibrary{arrows}
\usepackage{subcaption}
\DeclareMathOperator{\Tr}{Tr}

\newcommand{\ri}{\mathrm{i}}
\renewcommand{\th}{\theta}

\newcommand{\cob}{\delta}

\newcommand{\hf}{\frac{1}{2}}

\renewcommand{\b}[1]{\overline{#1}}
\newcommand{\del}{\partial}

\newcommand{\bra}{\langle}
\newcommand{\ket}{\rangle}
\newcommand{\la}{\lambda}

\newcommand{\bt}{\beta}

\newcommand{\al}{\alpha}
\newcommand{\om}{\omega}

\newcommand{\rt}[1]{\sqrt{#1}}

\begin{document}

\title{Spectral form factor and semi-circle law in the time direction}

\author{Kazumi Okuyama}

\affiliation{Department of Physics, 
Shinshu University,\\
3-1-1 Asahi, Matsumoto, Nagano 390-8621, Japan}

\emailAdd{kazumi@azusa.shinshu-u.ac.jp}

\abstract{
We study the time derivative of the connected part of
spectral form factor, which we call the slope of ramp, 
in Gaussian matrix model.
We find a closed formula of the slope of ramp at finite $N$
with non-zero inverse temperature.
Using this exact result, we confirm numerically that
the slope of ramp exhibits a semi-circle law as a function of time.}

\maketitle

\section{Introduction \label{sec:intro}}
After the seminal work \cite{Garcia-Garcia:2016mno,Cotler:2016fpe},
the spectral form factor is intensively studied as a diagnostic
of the quantum chaotic behavior of the Sachdev-Ye-Kitaev (SYK) model \cite{KitaevTalks,Sachdev,Maldacena:2016hyu}, which is a solvable example of the
holographic model of  a certain black hole in two-dimension.
At late times, the spectral form factor
of SYK model exhibits a structure of the so-called \textit{ramp}
and \textit{plateau},  and
it is well-approximated by the behavior of the Gaussian Unitary Ensemble 
(GUE) random matrix model 
when the number of fermions mod 8 is 2 or 6 \cite{you}\footnote{See also \cite{Gharibyan:2018jrp,Hunter-Jones:2017crg,Li:2017hdt,Kanazawa:2017dpd,Saad:2018bqo,Garcia-Garcia:2018ruf,Nosaka:2018iat,Garcia-Garcia:2017bkg} for the study of spectral form factor
in SYK model and its supersymmetric generalizations.}.

\begin{figure}[thb]
\centering
\includegraphics[width=10cm]{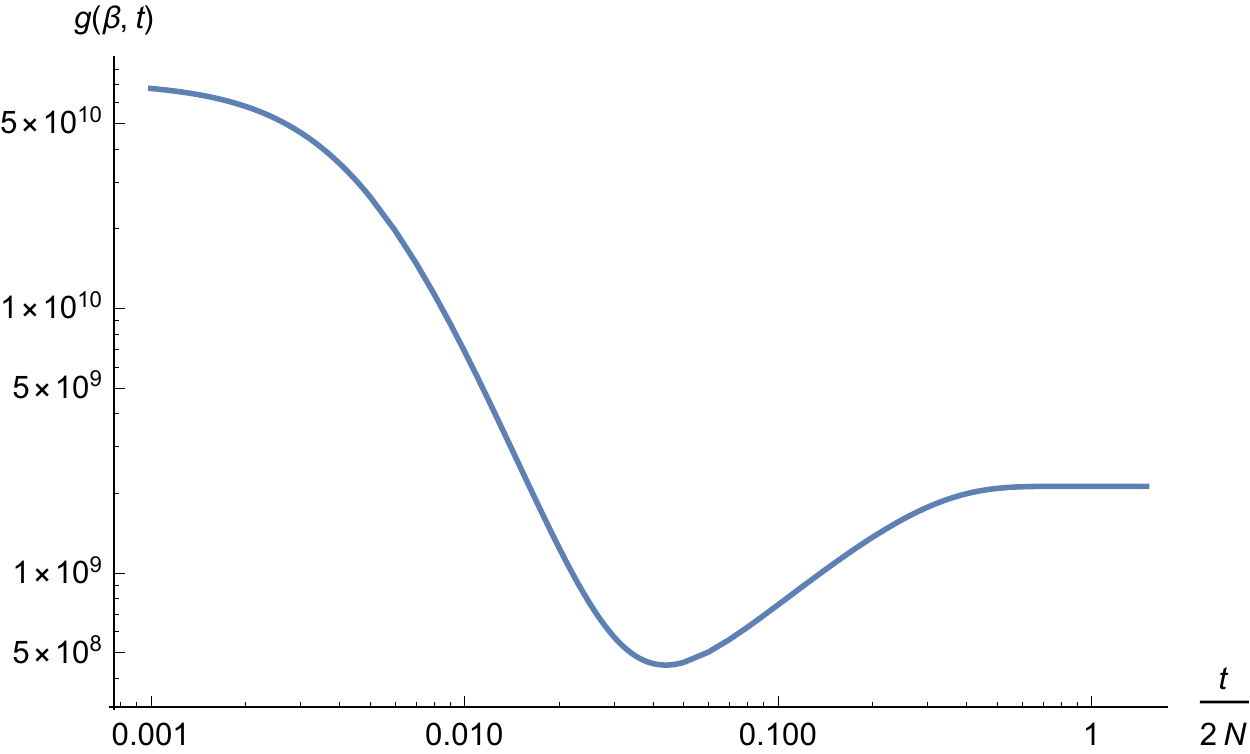}
\caption{Plot of the exact spectral form factor $g(\bt,t)$ 
in GUE for $\bt=5, N=500$.
}
  \label{fig:gtotal}
\end{figure}

In this paper, we will consider the  the 
spectral form factor $g(\bt,t)$ in GUE matrix model
with non-zero inverse temperature $\bt$.
We will show that
$g(\bt,t)$
is written exactly as a trace of 
an $N\times N$ matrix $A(z)$ defined in \eqref{eq:Amat}. 
$g(\bt,t)$ consists of two parts: the disconnected part
$g_{\text{disc}}(\bt,t)$ \eqref{eq:gdisc} and the connected part 
$g_{\text{conn}}(\bt,t)$ \eqref{eq:gconn}.
In Figure \ref{fig:gtotal}, we show the plot of this exact $g(\bt,t)$
for $\bt=5$ with the matrix size $N=500$.
As we can see from Figure \ref{fig:gtotal},
after the initial decay described by 
the disconnected part $g_{\text{disc}}(\bt,t)$, 
$g(\bt,t)$ has the structure of ramp
and plateau at late times. This late time behavior
comes from the connected part 
$g_{\text{conn}}(\bt,t)$ and it was studied 
extensively in the literature (see e.g. \cite{hikami,Liu:2018hlr} and references therein).
\footnote{The spectral form factor was first introduced in
\cite{jost} as a Fourier transform of the two-level correlation
function,
and it was observed that the spectral form factor exhibits a structure of dip,
which was originally called the ``correlation hole'' in \cite{jost}.}

The ramp is closely related to
the short range correlation of eigenvalues described by
the so-called sine kernel, and if we focus on the contribution from
a small window around some fixed eigenvalue the ramp grows linearly in $t$.
However, since $g(\bt,t)$ is defined by integrating over
the whole range of eigenvalue distribution, the actual ramp is not a linear function of $t$.

In this paper, we will study the non-linearity of ramp using the exact result at finite $N$.
To see the deviation from the linear behavior, it is natural
to consider the time derivative of $g_{\text{conn}}(\bt,t)$, which we will call the 
\textit{slope of ramp}.
If the ramp were a linear function of $t$, the slope of ramp would be a constant.
However, the actual slope of ramp is not constant in time.
It turns out that the slope of ramp obeys the semi-circle law as a function of time.
This is a direct consequence of the semi-circle law of eigenvalue distribution,
of course, but there is an interesting twist:
the slope of ramp corresponds to the eigenvalues and the time 
corresponds to
the eigenvalue density (see Figure \ref{fig:circle} for the detail
of this correspondence). In other words,
the eigenvalue density manifests itself as the time direction 
in the graph of the slope of ramp.

This paper is organized as follows.
In section \ref{sec:exact}, we write down the exact
closed form expression of the slope of ramp
$\del_t g_{\text{conn}}(\bt,t)$ at finite $N$.
In section \ref{sec:largeN}, we compute the late time behavior of
$g_{\text{conn}}(\bt,t)$ in the large $N$
limit. We point out that after an appropriate change of variable
\eqref{eq:sbt}, the slope of ramp obeys the semi-circle law as a function of time.
In section \ref{sec:plot}, we plot the slope of ramp as a function of time
using our exact result at finite $N$ for both $\bt=0$ and $\bt\ne0$ cases,
and confirm that the slope of ramp exhibits the semi-circle law. 
In section \ref{sec:smallt}, we consider the slope of ramp in the small $t$ regime.
Finally, we conclude in section \ref{sec:conclusion}.
In Appendix \ref{app:mat}, we explain how to compute 
$\Tr A(z)$ and $\Tr A(z_1)A(z_2)$.

\section{Exact slope of ramp at finite $N$ \label{sec:exact}}
In this paper we consider the spectral form factor in Gaussian
matrix model defined by
\begin{equation}
\begin{aligned}
 g(\bt,t)=\Bigl\bra \Tr e^{-(\bt+\ri t)H}
\Tr e^{-(\bt-\ri t)H}\Bigr\ket
&=\frac{\int dHe^{-\frac{N}{2}\Tr H^2}\Tr e^{-(\bt+\ri t)H}
\Tr e^{-(\bt-\ri t)H}}{\int dHe^{-\frac{N}{2}\Tr H^2}},
\end{aligned} 
\label{eq:def-g}
\end{equation}
where the integral is over the $N\times N$ hermitian matrix $H$.
By definition, $g(\bt,t)$ is an even function of $t$. Moreover,
since the Gaussian measure is invariant under $H\to -H$, 
$g(\bt,t)$ is independent of the sign
of $\bt$.
In the following we will assume that $\bt$ and $t$
are both positive without loss of generality:
\begin{equation}
\begin{aligned}
 \bt\geq0,\quad t\geq0.
\end{aligned} 
\end{equation}

In the normalization of Gaussian measure
in \eqref{eq:def-g},
the eigenvalue $\mu$ of matrix $H$ is distributed along the cut $\mu\in[-2,2]$ 
in the large $N$ limit,
and the eigenvalue density $\rho(\mu)$ is given by the
Wigner semi-circle law
\begin{equation}
\begin{aligned}
 \rho(\mu)=\frac{1}{2\pi}\rt{4-\mu^2} .
\end{aligned} 
\label{eq:wigner}
\end{equation}

As pointed out in \cite{delCampo:2017bzr},
$g(\bt,t)$ in \eqref{eq:def-g}
is formally equivalent to the correlator of 1/2 BPS Wilson loops in
4d $\mathcal{N}=4$ Super Yang-Mills (SYM) theory,
which is also given by the Gaussian matrix model via
the supersymmetric localization \cite{Erickson:2000af,Drukker:2000rr,Pestun:2007rz}.
Thus, we can immediately find the exact form of $g(\bt,t)$
by borrowing the known result of $\mathcal{N}=4$ SYM
in \cite{Drukker:2000rr,Kawamoto:2008gp,Okuyama:2018aij}.
To do this, it is convenient to rescale the matrix 
\begin{equation}
\begin{aligned}
 H=\rt{\frac{2}{N}}M,
\end{aligned} 
\end{equation}
so that the measure becomes $\int dM e^{-\Tr M^2}$.
In this normalization, $g(\bt,t)$ is written as
\begin{equation}
\begin{aligned}
 g(\bt,t)=\Bigl\bra \Tr e^{\frac{\bt+\ri t}{\rt{N}}\rt{2}M}
\Tr e^{\frac{\bt-\ri t}{\rt{N}}\rt{2}M}\Bigr\ket.
\end{aligned} 
\label{eq:gmat}
\end{equation}
On the other hand, the correlator of 1/2 BPS Wilson loops with winding number
$k_i$ is given by \cite{Okuyama:2018aij}
\begin{equation}
\begin{aligned}
 \left\bra\prod_i \Tr e^{k_i\rt{\frac{\la}{4N}}\rt{2}M}\right\ket,
\end{aligned} 
\label{eq:Wmat}
\end{equation} 
where $\la$ denotes the 't Hooft coupling of $\mathcal{N}=4$ SYM.
Comparing \eqref{eq:gmat} and \eqref{eq:Wmat}, we find a dictionary between
Wilson loops in $\mathcal{N}=4$ SYM and the spectral form factor
\begin{equation}
\begin{aligned}
 k_i\rt{\la}~\leftrightarrow~2(\bt\pm\ri t).
\end{aligned} 
\end{equation}

As shown in \cite{Fiol:2013hna,Okuyama:2018aij},
the correlator of $\Tr e^{z\rt{2}M}$ is written in terms of the $N\times N$
symmetric matrix
$A(z)$ defined by
\begin{equation}
\begin{aligned}
 A(z)_{i,j}=\rt{\frac{i!}{j!}}e^{\frac{z^2}{2}}z^{j-i}
L_i^{j-i}(-z^2), \quad(i,j=0,\cdots,N-1),
\end{aligned} 
\label{eq:Amat}
\end{equation}
where $L_n^\al(x)$ denotes the associated Laguerre polynomial.
The one-point function is given by
the trace of $A(z)$ (see Appendix \ref{app:mat} for a derivation of this result)
\begin{equation}
\begin{aligned}
 \Bigl\bra \Tr e^{z\rt{2}M}\Bigr\ket
=\Tr A(z)=e^{\frac{z^2}{2}}L_{N-1}^1(-z^2).
\end{aligned} 
\end{equation}

The spectral form factor $g(\bt,t)$ in \eqref{eq:gmat}
is a two-point function
of $\Tr e^{z\rt{2}M}$ and $\Tr e^{\b{z}\rt{2}M}$ with
\begin{equation}
\begin{aligned}
 z=\frac{\bt+\ri t}{\rt{N}},\quad
 \b{z}=\frac{\bt-\ri t}{\rt{N}}.
\end{aligned} 
\label{eq:z-bt}
\end{equation}
One can naturally decompose $g(\bt,t)$  into the disconnected part 
$g_{\text{disc}}(\bt,t)$
and the connected part $g_{\text{conn}}(\bt,t)$
\begin{equation}
\begin{aligned}
 g(\bt,t)= g_{\text{disc}}(\bt,t)+ g_{\text{conn}}(\bt,t).
\end{aligned} 
\end{equation}
The disconnected part is given by a product of one-point functions
\begin{equation}
\begin{aligned}
 g_{\text{disc}}(\bt,t)=\Tr A(z)\Tr A(\b{z})
=e^{\frac{z^2+\b{z}^2}{2}}L_{N-1}^1(-z^2)L_{N-1}^1(-\b{z}^2),
\end{aligned} 
\label{eq:gdisc}
\end{equation}
where $z$ and $\b{z}$ are defined in \eqref{eq:z-bt}.
This part is responsible for the early time decay of $g(\bt,t)$, 
which we will not consider in this paper.

The late time behavior of $g(\bt,t)$, the so-called ramp and plateau,
comes form the connected part. Using the result in 
\cite{Drukker:2000rr,Kawamoto:2008gp,Okuyama:2018aij},
$g_{\text{conn}}(\bt,t)$ is written as
\begin{equation}
\begin{aligned}
 g_{\text{conn}}(\bt,t)&=\Tr \Bigl[A(z+\b{z})-A(z)A(\b{z})\Bigr].
\end{aligned} 
\label{eq:gconn}
\end{equation}
Since $z+\b{z}=\frac{2\bt}{\rt{N}}$,
the first term of \eqref{eq:gconn} is independent of time
and it sets the value of plateau 
\begin{equation}
\begin{aligned}
 g_{\text{plateau}}(\bt)&=\Tr A(z+\b{z})
=e^{\frac{2\bt^2}{N}}L_{N-1}^1\Bigl(-\frac{4\bt^2}{N}\Bigr).
\end{aligned} 
\end{equation}
Using the result of Wilson loop in $\mathcal{N}=4$ SYM \cite{Erickson:2000af}, 
the large $N$ limit of $g_{\text{plateau}}(\bt)$
with fixed $\bt$ is given by\footnote{The initial value of the
disconnected part $g_{\text{disc}}(\bt,t=0)$ is order $N^2$
in the large $N$ limit
\begin{equation}
\begin{aligned}
 g_{\text{disc}}(\bt,t=0)\approx N^2\frac{I_1(2\bt)^2}{\bt^2}.
\end{aligned} 
\end{equation}
Note that this is larger than the value of plateau \eqref{eq:plateau-value}
by a factor of $N$.
}
\begin{equation}
\begin{aligned}
  g_{\text{plateau}}(\bt)\approx N\frac{I_1(4\bt)}{2\bt},
\end{aligned} 
\label{eq:plateau-value}
\end{equation}
where $I_n(x)$ denotes the modified Bessel function of the first kind.

The non-trivial time dependence comes from the second term of \eqref{eq:gconn} 
\begin{equation}
\begin{aligned}
 g_{\text{ramp}}(\bt,t)&=-\Tr \Bigl[A(z)A(\b{z})\Bigr].
\end{aligned} 
\end{equation}
In what follows, we will consider the time derivative of $g_{\text{ramp}}(\bt,t)$,
which we call the \textit{slope of ramp}.
Since $g_{\text{plateau}}(\bt)$ is independent of time, 
the slope of ramp is equal to the time derivative of the connected part
of spectral form factor
\begin{equation}
\begin{aligned}
 \frac{\del g_{\text{ramp}}}{\del t}(\bt,t)=\frac{\del g_{\text{conn}}}{\del t}(\bt,t).
\end{aligned} 
\end{equation}

As explained in Appendix \ref{app:mat},
we can write down a closed form expression of the slope of ramp
\begin{equation}
\begin{aligned}
 \frac{\del g_{\text{conn}}}{\del t}(\bt,t)&=
\frac{N}{\bt}e^{\frac{z^2+\b{z}^2}{2}}\text{Im}
\Bigl[L_N(-z^2)L_{N-1}(-\b{z}^2)\Bigr] .
\end{aligned} 
\label{eq:slope-bt}
\end{equation}
By taking the limit $\bt\to0$ of \eqref{eq:slope-bt}, the slope of ramp for $\bt=0$ becomes
\begin{equation}
\begin{aligned}
 \frac{\del g_{\text{conn}}}{\del t}(0,t)=
2te^{-\frac{t^2}{N}}\Biggl[L_{N-1}\Bigl(\frac{t^2}{N}\Bigr)
L_{N-1}^1\Bigl(\frac{t^2}{N}\Bigr)-L_{N}\Bigl(\frac{t^2}{N}\Bigr)
L_{N-2}^1\Bigl(\frac{t^2}{N}\Bigr)\Biggr] .
\end{aligned} 
\label{eq:slope-0}
\end{equation}

We are interested in the large $N$ limit of the slope of ramp
\eqref{eq:slope-bt} and  
\eqref{eq:slope-0}.
When $\bt=0$, as pointed out in \cite{hikami},
$\del_t g_{\text{conn}}(0,t)$ in \eqref{eq:slope-0}
happens to be equal to the eigenvalue
density in the Wishart-Laguerre ensemble, which is known to 
obey the semi-circle law in the large $N$ limit.\footnote{See 
eq.(3.16) and eq.(3.30) in \cite{Brezin:1995dp} (see also \cite{Verbaarschot:1993pm}).
The eigenvalue density of Wishart-Laguerre ensemble
$\rho(\mu)=\mu\tilde{\rho}(\mu^2)$ in \cite{Brezin:1995dp} is equal to
$\hf\del_t g_{\text{conn}}(0,t)$ under the identification $\mu=t/N$;
eq.(3.30) in \cite{Brezin:1995dp} corresponds to the exact finite $N$ result
of $\del_t g_{\text{conn}}(0,t)$ in \eqref{eq:slope-0}, while eq.(3.16)
in \cite{Brezin:1995dp} represents its large $N$ limit.} 
However, the large $N$ limit of $\del_t g_{\text{conn}}(\bt,t)$ 
with non-zero $\bt$ is not well studied in the literature.
In section \ref{sec:largeN},
we will numerically study the large $N$ behavior of the exact result \eqref{eq:slope-bt} and  
\eqref{eq:slope-0}.

Before doing this numerical study, in the next section we will review the 
analytic derivation of the large $N$ behavior of ramp
in \cite{hikami,Liu:2018hlr}.

\section{Large $N$ limit of the slope of ramp \label{sec:largeN}}
The large $N$ limit of $g_{\text{conn}}(\bt,t)$
is written in terms of the connected part of the two-level
correlation function $\rho^{(2)}(\mu_1,\mu_2)$
\begin{equation}
\begin{aligned}
 g_{\text{conn}}(\bt,t)&=\int d\mu_1d\mu_2\rho^{(2)}(\mu_1,\mu_2)
e^{(\bt+\ri t)\mu_1}e^{(\bt-\ri t)\mu_2}\\
&=\int d\mu_1d\mu_2\rho^{(2)}(\mu_1,\mu_2)e^{\ri t(\mu_1-\mu_2)+\bt(\mu_1+\mu_2)}.
\end{aligned} 
\end{equation}
At late times $t\gg1$, the dominant contribution comes from
the region $|\mu_1-\mu_2|\ll1$.
Thus we can use the universal form of the short range correlation, known as the \textit{sine kernel}
(see e.g. \cite{Mehta})
\begin{equation}
\begin{aligned}
g_{\text{conn}}(\bt,t) \approx
-N^2\int d\mu_1d\mu_2\left[ \frac{\sin N\pi (\mu_1-\mu_2) \rho\bigl(\frac{\mu_1+\mu_2}{2}\bigr)}
{N\pi (\mu_1-\mu_2)}\right]^2 e^{\ri t(\mu_1-\mu_2)+\bt (\mu_1+\mu_2)}.
\end{aligned} 
\label{eq:gconn-sine}
\end{equation}
Introducing the variables $\om$ and $u$  by
\begin{equation}
\begin{aligned}
 \om=2N(\mu_1-\mu_2),\quad
u=\frac{\mu_1+\mu_2}{4},
\end{aligned} 
\end{equation}
\eqref{eq:gconn-sine} is rewritten as
\begin{equation}
\begin{aligned}
g_{\text{conn}}(\bt,t) \approx
-\frac{4N}{\pi^2}\int dud\om \frac{\sin^2  \frac{\pi}{2} \rho(2u)\om}{\om^2} e^{\ri \om\tau+4\bt u},
\end{aligned} 
\label{eq:g-omint}
\end{equation}
where $\tau$ is given by
\begin{equation}
\begin{aligned}
 \tau=\frac{t}{2N} .
\end{aligned} 
\label{eq:tau-def}
\end{equation}
In the large $N$ limit, the integration region of $\om$ can be extended to $\om\in[-\infty,\infty]$, 
and the $\om$-integral is explicitly evaluated as \cite{hikami}
\begin{equation}
\begin{aligned}
 \int_{-\infty}^\infty d\om\frac{\sin^2 \frac{\pi}{2} \rho(2u)\om}{\om^2}e^{\ri \om\tau}
=\left\{
\begin{aligned}
&\frac{\pi}{2} \big(\pi\rho(2u)-\tau\big),\quad & (\pi\rho(2u)>\tau),\\
&0, \quad & (\pi\rho(2u)<\tau).
\end{aligned}
\right.
\end{aligned} 
\label{eq:relu}
\end{equation}
The condition $\pi\rho(2u)>\tau$ limits the range of $u$-integration
to $u\in[-u_\tau,u_\tau]$, where $u_\tau$ is determined by $\pi\rho(2u_\tau)=\tau$.
From the explicit form of eigenvalue density in \eqref{eq:wigner},  
we find
\begin{equation}
\begin{aligned}
\pi\rho(2u_\tau)= \rt{1-u_\tau^2}=\tau,
\end{aligned} 
\label{eq:u-tau}
\end{equation}
and $u_\tau$ is given by
\begin{equation}
\begin{aligned}
 u_\tau=\rt{1-\tau^2}.
\end{aligned} 
\label{eq:utau-sqrt}
\end{equation}
Since the maximal value of $\pi\rho(2u_\tau)$ is one,
$\tau=1$ is the critical value at which
the behavior of $g_{\text{conn}}(\bt,t)$ changes discontinuously from ramp to plateau.
In the following, we will
consider the ramp regime $\tau<1$.
When $\tau<1$, plugging \eqref{eq:relu}
into \eqref{eq:g-omint} we find that $g_{\text{conn}}(\bt,t)$ is written as
\begin{equation}
\begin{aligned}
 g_{\text{conn}}(\bt,t)=\frac{2N}{\pi}\int_{-u_\tau}^{u_\tau}du \,e^{4\bt u}\bigl(
\tau-\pi \rho(2u)\bigr).
\end{aligned} 
\label{eq:gconn-uint}
\end{equation}
Let us consider the time derivative of
$g_{\text{conn}}(\bt,t)$ in \eqref{eq:gconn-uint}. 
The $t$-derivative of the boundary term $\pm u_\tau$ vanishes
due to the condition \eqref{eq:u-tau}.
Thus, the $t$-derivative of \eqref{eq:gconn-uint} comes only from the
derivative of integrand 
\begin{equation}
\begin{aligned}
 \frac{\del g_{\text{conn}}}{\del t}(\bt,t)&=\frac{2N}{\pi}\int_{-u_\tau}^{u_\tau}du\, e^{4\bt u}\frac{\del\tau}{\del t}=
\frac{1}{\pi}\int_{-u_\tau}^{u_\tau}du\, e^{4\bt u}=
\frac{\sinh 4\bt u_\tau}{2\pi\bt}.
\end{aligned} 
\label{eq:delg-sinh}
\end{equation}

Let us take a closer look at the case of $\bt=0$.
By setting $\bt=0$ in \eqref{eq:delg-sinh}, one can see
that $\del_t g_{\text{conn}}(0,t)$ is proportional
to $u_\tau$
\begin{equation}
\begin{aligned}
 \frac{\del g_{\text{conn}}}{\del t}(0,t)=\frac{2}{\pi}u_\tau.
\end{aligned} 
\end{equation}
Introducing the rescaled slope of ramp $s(0,t)$ by 
\begin{equation}
\begin{aligned}
 s(0,t):=\frac{\pi}{2}\frac{\del g_{\text{conn}}}{\del t}(0,t)=u_\tau,
\end{aligned} 
\label{eq:s0-def}
\end{equation}
it follows from \eqref{eq:utau-sqrt} that $s(0,t)$ obeys the semi-circle law
\begin{equation}
\begin{aligned}
 s(0,t)^2+\tau^2=1.
\end{aligned} 
\label{eq:st-circle}
\end{equation}

\begin{figure}[thb]
\centering
\begin{tikzpicture}
\draw[->,thick] (-5.5,0)--(5.5,0);
\draw[->,thick] (0,-0.5)--(0,5.8);
\coordinate (u) at (5.5,0) node at (u) [right] {$u$};
\coordinate (r) at (0,5.8) node at (r) [above] {$\pi\rho(2u)$};
\draw[thick,blue] (5,0) arc (0:180:5);
\coordinate (up) at (4,0) node at (up) [below] {$u_\tau$};
\coordinate (um) at (-4,0) node at (um) [below] {$-u_\tau$};
\coordinate (O) at (0,-0.22) node at (O) [left]{$0$};
\coordinate (t) at (0,3.22) node at (t) [left] {$\tau$};
\coordinate (s) at (2,3) node at (s) [below] {$s(\bt,t)$};
\coordinate (c1) at (5,0) node at (c1) [below] {$1$};
\coordinate (c2) at (-5,0) node at (c2) [below] {$-1$};
\coordinate (c3) at (0,5.22) node at (c3) [left] {$1$};
\draw[thick,red, dotted] (-4,3)--(0,3);
\draw[<->,thick,red] (4,3)--(0,3);
\draw[thick,dashed] (-4,3)--(-4,0);
\draw[thick,dashed] (4,3)--(4,0);
\end{tikzpicture}
\caption{This figure shows the interpretation of $\tau$ and $s(\bt,t)$
in the eigenvalue distribution. The blue semi-circle is the graph of eigenvalue density $\pi\rho(2u)=\rt{1-u^2}$. The time slice $\pi\rho(2u)=\tau$ is represented by
the horizontal red line. The \textit{slope of ramp} $s(\bt,t)=u_\tau$ 
corresponds to the length of solid red line.} 
\label{fig:circle}
\end{figure}
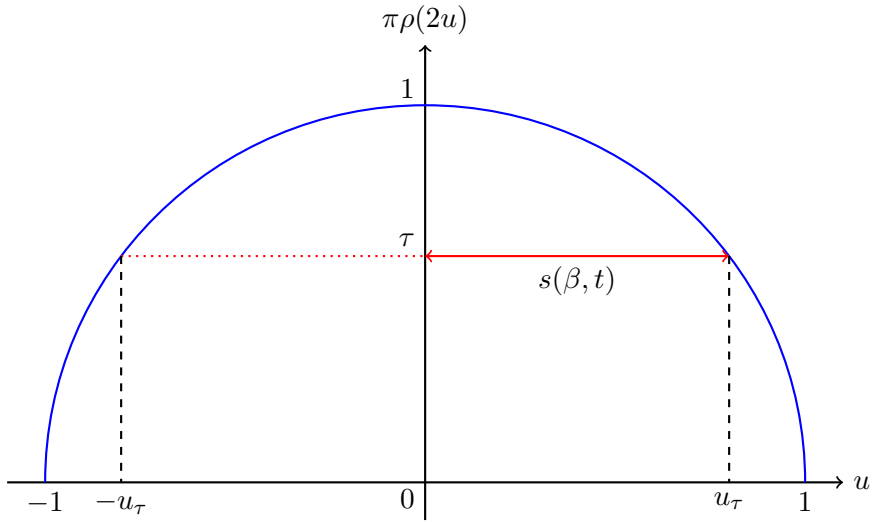

When $\bt\ne0$, 
one can similarly define the quantity $s(\bt,t)$
by applying
the inverse function of sinh to $\del_t g_{\text{conn}}$ in \eqref{eq:delg-sinh}: 
\begin{equation}
\begin{aligned}
 s(\bt,t):=\frac{1}{4\bt}\text{arcsinh}\left(2\pi\bt \frac{\del g_{\text{conn}}}{\del t}(\bt,t)\right)
=u_\tau.
\end{aligned} 
\label{eq:sbt}
\end{equation}
Again, from \eqref{eq:utau-sqrt} it follows that 
$s(\bt,t)$ obeys the semi-circle law
\begin{equation}
\begin{aligned}
s(\bt,t)^2+\tau^2=1 .
\end{aligned} 
\label{eq:sb-circle}
\end{equation}
In the rest of this paper, we will use the name ``slope of ramp''
for both $\del_t g_{\text{conn}}(\bt,t)$ and $s(\bt,t)$ interchangeably.

In Figure \ref{fig:circle}, we show the interpretation of $s(\bt,t)$
in the Wigner semi-circle distribution.
Here we comment on some feature of this figure:
\begin{itemize}
\item The time $\tau$ corresponds to the vertical axis in Figure \ref{fig:circle}.
Namely, $\tau$ probes the value of eigenvalue density (see \eqref{eq:u-tau}).
\item The \textit{slope of ramp} $s(\bt,t)$ in \eqref{eq:sbt}
corresponds to the horizontal direction in Figure \ref{fig:circle}.
In other words, $s(\bt,t)$ plays the role of eigenvalue.
\item The point $(s(\bt,t),\tau)$ lies on the unit semi-circle \eqref{eq:sb-circle}.
\end{itemize}

Before closing this section, we note in passing that
the large $N$ limit of $g_{\text{conn}}(\bt,t)$ is easily obtained by integrating 
$\del_t g_{\text{conn}}$ in \eqref{eq:delg-sinh}
\begin{equation}
\begin{aligned}
 g_{\text{conn}}(\bt,t)
=g_{\text{conn}}(\bt,0)+2N\int_0^\tau d\tau' \frac{\sinh4\bt\rt{1-\tau'^2}}{2\pi \bt}.
\end{aligned} 
\end{equation}
After a change of variable $\tau=\sin\th$,
this integral can be performed by using the relation
\begin{equation}
\begin{aligned}
 \sinh(4\bt\cos\th)=2\sum_{n=1}^\infty I_{2n-1}(4\bt)\cos(2n-1)\th.
\end{aligned} 
\end{equation}
Then we find
\begin{equation}
\begin{aligned}
 g_{\text{conn}}(\bt,t)=
g_{\text{conn}}(\bt,0)+
\frac{N}{\pi\bt}\left[
I_1(4\bt)\th
+\sum_{n=1}^\infty \frac{I_{2n+1}(4\bt)+
I_{2n-1}(4\bt)}{2n}\sin 2n\th\right],
\end{aligned} 
\label{eq:gconn-Ibt}
\end{equation}
where $\th$ is related to time $\tau$ by
\begin{equation}
\begin{aligned}
 \th=\arcsin(\tau).
\end{aligned} 
\end{equation}
Note that the initial value $g_{\text{conn}}(\bt,0)$ is given by
\begin{equation}
\begin{aligned}
 g_{\text{conn}}(\bt,0)=
\Tr\left[A\Bigl(\frac{2\bt}{\rt{N}}\Bigr)-A\Bigl(\frac{\bt}{\rt{N}}\Bigr)^2\right].
\end{aligned} 
\label{eq:gc-init}
\end{equation}
When $\bt=0$ this initial value vanishes $g_{\text{conn}}(0,0)=0$, but
it is non-zero for $\bt\ne0$. The large $N$ limit of $g_{\text{conn}}(\bt,0)$
in \eqref{eq:gc-init} can be obtained by
borrowing the result of two-point correlator of 1/2 BPS Wilson loops in $\mathcal{N}=4$
SYM \cite{Akemann:2001st,Giombi:2009ms,Okuyama:2018aij}
\begin{equation}
\begin{aligned}
 g_{\text{conn}}(\bt,0)= \bt I_0(2\bt)I_1(2\bt)+\mathcal{O}(N^{-2}).
\end{aligned} 
\end{equation}

When $\bt=0$, \eqref{eq:gconn-Ibt} reproduces the known result
in \cite{hikami,Liu:2018hlr}
\begin{equation}
\begin{aligned}
 g_{\text{conn}}(0,t)&=\frac{2N}{\pi}\Bigl(\th+\hf\sin2\th\Bigr)
=\frac{2N}{\pi}\Bigl(\arcsin(\tau)+\tau\rt{1-\tau^2}\Bigr).
\end{aligned} 
\end{equation}
We have also checked that the small $\bt$ expansion of our result \eqref{eq:gconn-Ibt}
is consistent with the $\mathcal{O}(\bt^2)$ term of $g_{\text{conn}}(\bt,t)$ 
computed in
\cite{Liu:2018hlr}.
\section{Plot of the exact slope of ramp \label{sec:plot}}
In this section, we will study numerically 
the behavior of the exact slope of ramp $s(\bt,t)$ at finite $N$.
Plugging the exact result of $\del_t g_{\text{conn}}(\bt,t)$ \eqref{eq:slope-bt}
into \eqref{eq:sbt}, we find the exact form of
$s(\bt,t)$ at finite $N$
\begin{equation}
\begin{aligned}
 s(\bt,t)= \frac{1}{4\bt}\text{arcsinh}\left(
2\pi Ne^{\frac{\bt^2-t^2}{N}}\text{Im}\Biggl[L_N\Bigl(-\frac{(\bt+\ri t)^2}{N}\Bigr)
L_{N-1}\Bigl(-\frac{(\bt-\ri t)^2}{N}\Bigr)\Biggr]\right).
\end{aligned} 
\label{eq:sbtexact} 
\end{equation}
When $\bt=0$, using the result of $\del_t g_{\text{conn}}(0,t)$ in \eqref{eq:slope-0}
the exact form of
$s(0,t)$ at finite $N$ becomes
\begin{equation}
\begin{aligned}
 s(0,t)=\pi 
te^{-\frac{t^2}{N}}\Biggl[L_{N-1}\Bigl(\frac{t^2}{N}\Bigr)
L_{N-1}^1\Bigl(\frac{t^2}{N}\Bigr)-L_{N}\Bigl(\frac{t^2}{N}\Bigr)
L_{N-2}^1\Bigl(\frac{t^2}{N}\Bigr)\Biggr].
\end{aligned}
\label{eq:s0exact} 
\end{equation}

\begin{figure}[htb]
\centering
\subcaptionbox{$s(0,t)$\label{sfig:sbt0}}{\includegraphics[width=7cm]{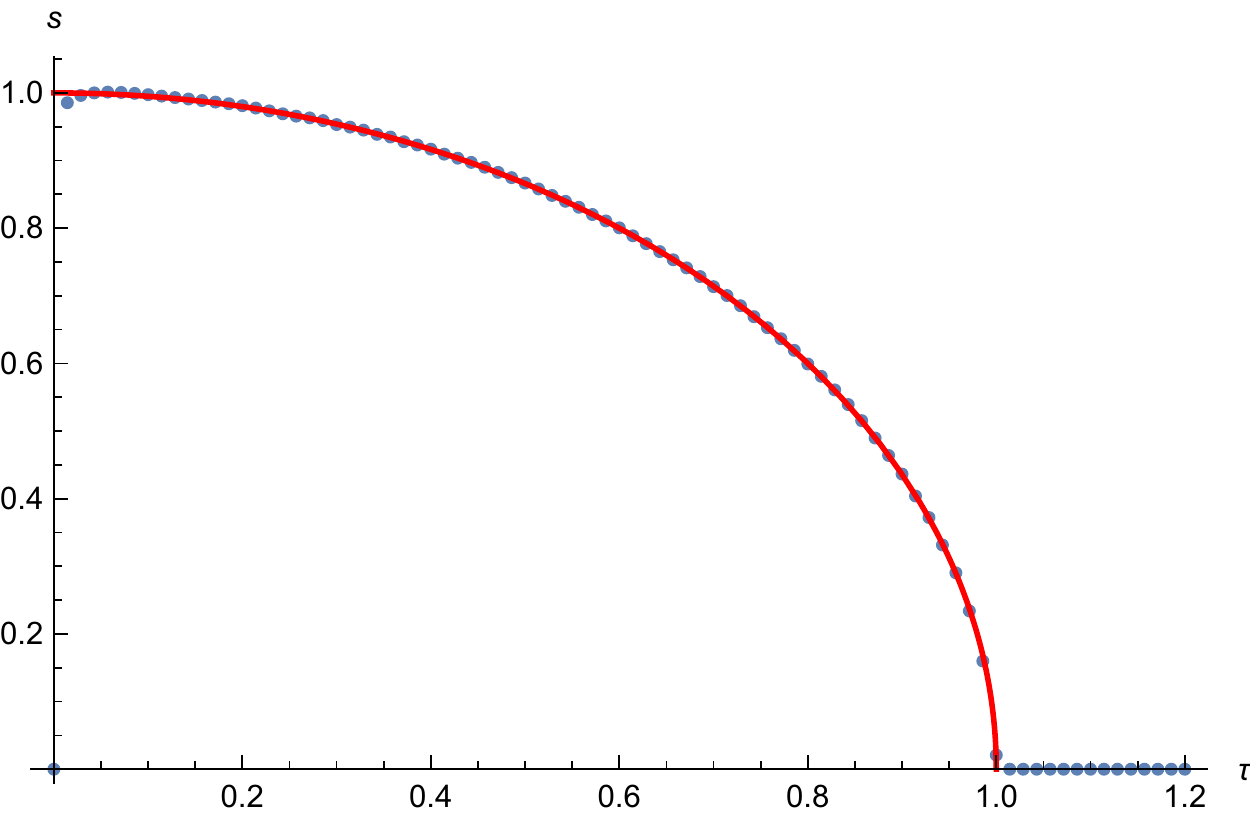}}
\hskip5mm
\subcaptionbox{$s(5,t)$\label{sfig:sbt5}}{\includegraphics[width=7cm]{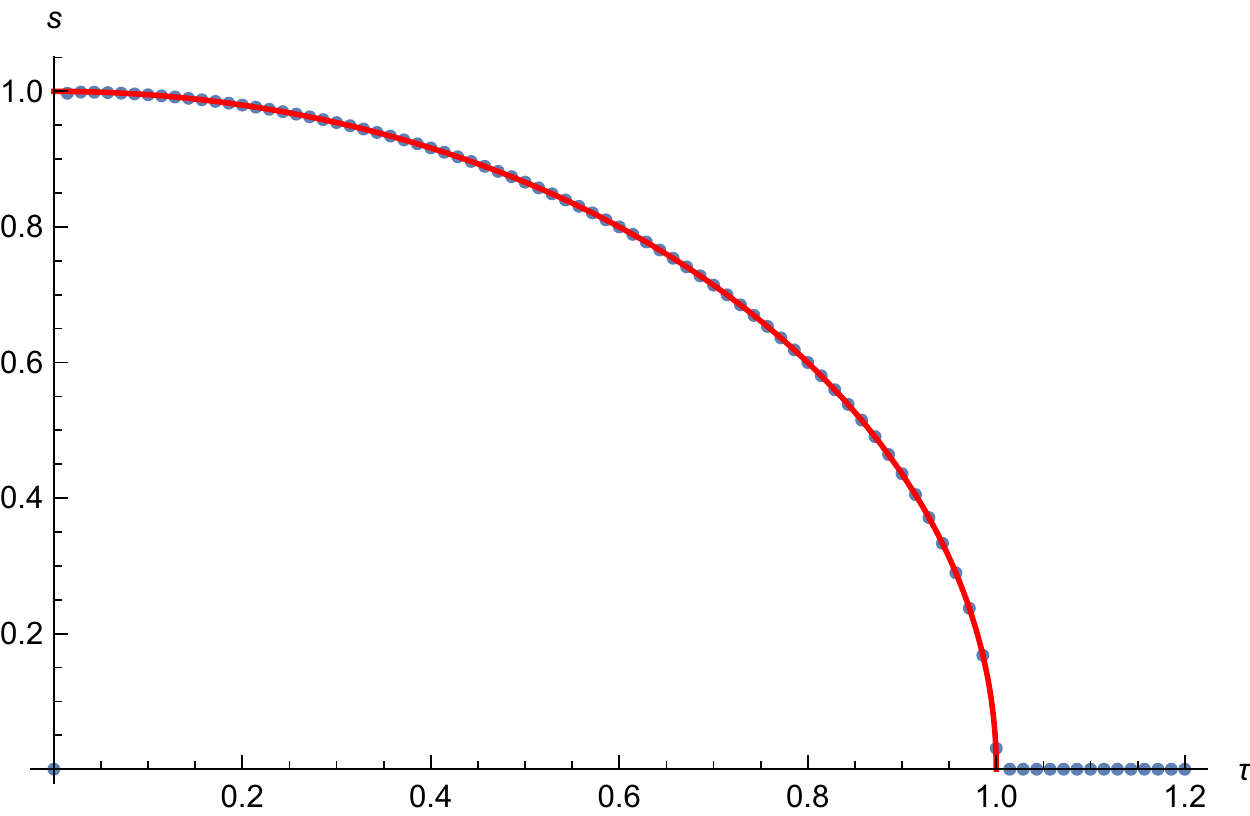}}
  \caption{Plot of $s(\bt,t)$
for \subref{sfig:sbt0} $\bt=0$ and \subref{sfig:sbt5} $\bt=5$.
The horizontal axis is the rescaled time $\tau=t/2N$.
The blue dots are the exact values of $s(\bt,t)$ at $N=500$ while the
red curve represents the semi-circle law $s=\rt{1-\tau^2}$.
}
  \label{fig:sbt}
\end{figure}

In Figure \ref{fig:sbt}, we plot the exact slope of ramp
$s(\bt,t)$ at $N=500$
as a function of  time $\tau=t/2N$.
One can clearly see that $s(\bt,t)$ obeys the semi-circle law as predicted by the large $N$
analysis in the previous section.
We emphasize that $s(\bt,t)$ is independent of $\bt$
in the large $N$ limit and it obeys the semi-circle law for both $\bt=0$ and $\bt\ne0$
as shown in \eqref{eq:st-circle} and \eqref{eq:sb-circle}.
On the other hand, $g_{\text{conn}}(\bt,t)$ itself has a non-trivial $\bt$-dependence,
whose explicit form in the large $N$ limit is given by \eqref{eq:gconn-Ibt}.

Note that the vertical and horizontal axes in Figure \ref{fig:circle}
are flipped in Figure \ref{fig:sbt}.
As we explained in the previous section, the
$\tau$-axis corresponds to the eigenvalue density and 
the $s$-axis
corresponds to the eigenvalues.
In other words, the eigenvalue density manifests itself as the time direction
in Figure \ref{fig:sbt}.

As we can see from Figure \ref{fig:sbt}, the slope of ramp vanishes beyond
the critical value $\tau=1$, which corresponds to the so-called Heisenberg time 
$t_H=2N$
where the plateau regime sets in.
This critical time is determined by the maximal value of
the eigenvalue density.

\section{Small $t$ behavior of the slope of ramp \label{sec:smallt}}

In this section we will consider the small $t$
behavior of the slope of ramp $s(\bt,t)$. Since $s(\bt,t)$ is an odd function of $t$,
its Taylor expansion starts from the linear term in $t$ \footnote{In \cite{Cotler:2017jue} it was observed numerically
that in the small $t$ regime
$g_{\text{conn}}(0,t)$ behaves as $g_{\text{conn}}(0,t)\sim t^2$.
This behavior simply follows from the fact that
$g_{\text{conn}}(0,t)$ is an even function of $t$ with
the initial value $g_{\text{conn}}(0,0)=0$, hence its Taylor expansion
starts from $t^2$.}.
From the exact result of $s(\bt,t)$ at finite $N$ in \eqref{eq:sbtexact},
we can compute the coefficient of this linear term
\begin{equation}
\begin{aligned}
s(\bt,t)= 
\pi e^{\frac{\bt^2}{N}}\left[
L_{N-1}\Bigl(-\frac{\bt^2}{N}\Bigr)
L_{N-1}^1\Bigl(-\frac{\bt^2}{N}\Bigr)-
L_{N}\Bigl(-\frac{\bt^2}{N}\Bigr)
L_{N-2}^1\Bigl(-\frac{\bt^2}{N}\Bigr)\right]t+\mathcal{O}(t^3).
\end{aligned} 
\end{equation}
In the large $N$ limit this becomes
\begin{equation}
\begin{aligned}
 s(\bt,t)=\pi \Bigl[I_0(2\bt)^2-I_1(2\bt)^2\Bigr]t+\mathcal{O}(t^3).
\end{aligned} 
\end{equation}
One can in principle compute the coefficient of $t^3,t^5,\cdots,$
as a function of $\bt$ using the 
exact result in \eqref{eq:sbtexact}.
However, the computation for general $\bt$ 
becomes tedious when we go to higher order terms.

Instead, here we focus on the  $\bt=0$ case
where the higher order coefficients are easily extracted from
the exact result at finite $N$ in \eqref{eq:s0exact}
\begin{equation}
\begin{aligned}
s(0,t)= \frac{\pi}{2}\left[ 2 t-2 t^3+t^5 +
\left(-\frac{5}{18}-\frac{1}{18N^2}\right)t^7+\mathcal{O}(t^9)\right].
\end{aligned} 
\label{eq:s0taylor}
\end{equation}
This expansion is valid until the first and the second terms in \eqref{eq:s0taylor} 
become comparable. The order of this time scale is 
\begin{equation}
\begin{aligned}
 t\sim \mathcal{O}(N^0).
\end{aligned} 
\end{equation}
Summing over the order $N^0$ terms in \eqref{eq:s0taylor}, we find
that  the large $N$ limit of $s(0,t)$ in the small $t$ regime
is given by the Bessel function
\begin{equation}
\begin{aligned}
 s(0,t)= \pi t\Bigl[J_0(2t)^2+J_1(2t)^2\Bigr]+\mathcal{O}(N^{-2}).
\end{aligned} 
\label{eq:bessel-osc}
\end{equation}

\begin{figure}[htb]
\centering
\includegraphics[width=10cm]{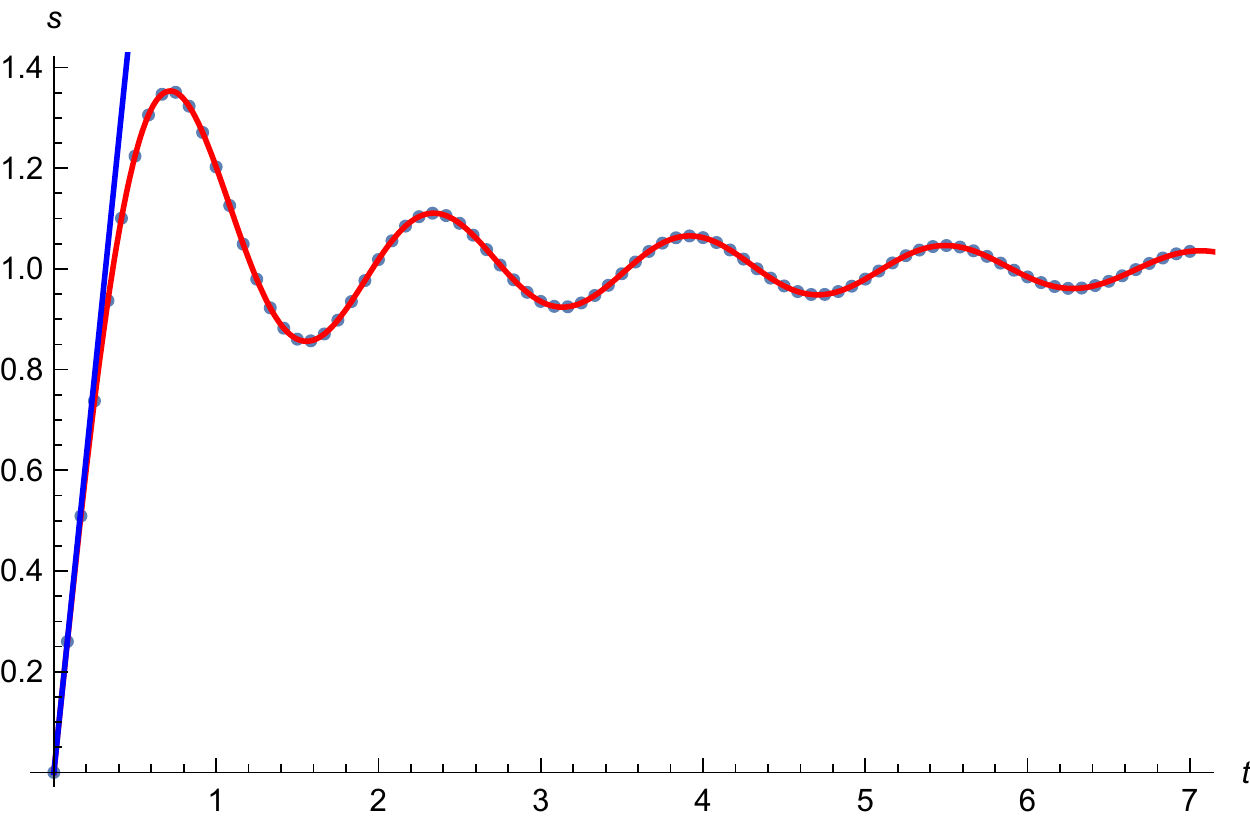}
\caption{Plot of $s(\bt=0,t)$
 in the small $t$ region.
The dots are the exact values of $s(0,t)$ for $N=500$.
The blue line is the first term $s=\pi t$
in the Taylor expansion of $s(0,t)$ in \eqref{eq:s0taylor},
while the red curve represents the Bessel function in
\eqref{eq:bessel-osc}.
This figure is a closeup of the small $t$ region of Figure \ref{sfig:sbt0}.
}
  \label{fig:smallt}
\end{figure}

In Figure \ref{fig:smallt}, we plot the exact $s(0,t)$ at $N=500$
in the small $t$ region. $s(0,t)$ grows linearly at very early time and then 
starts to oscillate around $s=1$. The linear behavior of
$s(0,t)$ around $t=0$ comes from the first term in the
Taylor expansion \eqref{eq:s0taylor}, while the  oscillating behavior
is captured by the Bessel function \eqref{eq:bessel-osc}
as discussed in \cite{hikami}.

When $t$ becomes of order $N$, the expression \eqref{eq:bessel-osc}
is no longer valid; 
$s(0,t)$ is described instead by the semi-circle law \eqref{eq:st-circle} when $t\sim \mathcal{O}(N)$.

\section{Conclusion \label{sec:conclusion}}
In this paper, we have studied the slope of ramp $s(\bt,t)$,
which is related to $\del_t g_{\text{conn}}(\bt,t)$ by \eqref{eq:sbt},
in the Gaussian matrix model.
We found the exact closed form expression of $s(\bt,t)$ in \eqref{eq:sbtexact} and
confirmed numerically that $s(\bt,t)$
obeys the semi-circle law as a function of time for both $\bt=0$ and $\bt\ne0$ cases.
Interestingly, in the plot of $s(\bt,t)$ the time direction plays the role of eigenvalue
density.

There are many interesting open questions. We list several avenues for
future research. 
The relation between $g_{\text{conn}}$ and the eigenvalue density
$\rho(\mu)$ in \eqref{eq:gconn-sine} is expected to be quite universal, and 
hence it is not restricted to the Gaussian matrix model.
It would be very interesting to study the slope of ramp in other models,
such as the SYK model,
and see if the eigenvalue density
manifests itself in the time direction for other models as well\footnote{See
\cite{Gaikwad:2017odv} for a study of
spectral form factor in hermitian matrix model with
a non-Gaussian potential.}.

It would be also interesting to generalize our study to the
higher point correlation function of $\Tr e^{-(\bt\pm\ri t)H}$.
In the case of Gaussian matrix model, the exact form of the connected part of
higher point function
was recently studied in \cite{Okuyama:2018aij}.
It would be interesting to see if the multi-point correlator of
eigenvalues $\rho^{(n)}(\mu_1,\cdots,\mu_n)$
appears in the time dependence of higher point functions of $\Tr e^{-(\bt\pm\ri t)H}$
in the large $N$ limit. To see this, we need to go beyond the ``box approximation''
used in \cite{Liu:2018hlr}.

\acknowledgments
I would like to thank Nick Hunter-Jones for correspondence
and careful reading of the manuscript.
This work was supported in part by JSPS KAKENHI Grant Number 16K05316.

\appendix
\section{Computation of $\Tr A(z)$ and $\Tr A(z_1)A(z_2)$ \label{app:mat}}
As discussed in \cite{Okuyama:2018aij}, the correlator
of $\Tr e^{z\rt{2}M}$ in Gaussian matrix model
with measure $\int dM e^{-\Tr M^2}$ is easily computed 
by using the harmonic oscillator
\begin{equation}
\begin{aligned}
 {[a,a^\dagger]}=1,
\end{aligned} 
\end{equation}
which is basically equivalent to the method of orthogonal polynomials
for solving hermitian matrix models. 
The result is written in terms of the $N\times N$ symmetric matrix
$A(z)$ with matrix element
\begin{equation}
\begin{aligned}
 A(z)_{i,j}=
A(z)_{j,i}=\bra i|e^{z(a+a^\dagger)}|j\ket,\quad (i,j=0,\cdots,N-1),
\end{aligned}
\label{eq:Aij} 
\end{equation}
where $|i\ket$ is the orthonormal basis
\begin{equation}
\begin{aligned}
 |i\ket=\frac{(a^\dagger)^i}{\rt{i!}}|0\ket, \quad \bra i|j\ket=\cob_{i,j}.
\end{aligned} 
\end{equation}
Using the generating function of Laguerre polynomial
\begin{equation}
\begin{aligned}
 (1+t)^\al e^{xt}=\sum_{n=0}^\infty L^{\al-n}_n(-x)t^n,
\end{aligned} 
\end{equation}
one can evaluate the matrix element in \eqref{eq:Aij} as
\begin{equation}
\begin{aligned}
 A(z)_{i,j}&=
e^{-\frac{z^2}{2}}\bra i|e^{za}e^{za^\dagger}|j\ket
=\frac{1}{\rt{i!j!}}\del_s^i\del_t^j e^{-\frac{z^2}{2}}
\bra 0|e^{(s+z)a}e^{(t+z)a^\dagger}|0\ket\Bigr|_{s=t=0}\\
&=\frac{1}{\rt{i!j!}}\del_s^i\del_t^j e^{-\frac{z^2}{2}}e^{(s+z)(t+z)}\Bigr|_{s=t=0}
=\frac{1}{\rt{i!j!}}\del_s^i e^{\frac{z^2}{2}+zs}(s+z)^j\Bigr|_{s=0}\\
&=\rt{\frac{i!}{j!}}e^{\frac{z^2}{2}}z^{j-i}
L_i^{j-i}(-z^2).
\end{aligned} 
\end{equation}

Let us first consider the trace of $A(z)$
\begin{equation}
\begin{aligned}
 \Tr A(z)=\sum_{i=0}^{N-1}A(z)_{i,i}=\sum_{i=0}^{N-1}\bra i|e^{z(a+a^\dagger)}|i\ket.
\end{aligned} 
\end{equation}
To evaluate this trace, it is convenient to
rewrite this as a trace in the total Hilbert space $\mathcal{H}$
of harmonic oscillator
\begin{equation}
\begin{aligned}
 \Tr A(z)=\Tr_\mathcal{H} \bigl(e^{z(a+a^\dagger)}P\bigr),
\end{aligned} 
\label{eq:TrA-Hilb}
\end{equation}
where $P$ denotes the projector to the first $N$ states
\begin{equation}
\begin{aligned}
 P=\sum_{i=0}^{N-1}|i\ket\bra i|,
\end{aligned} 
\end{equation}
and $\Tr_{\mathcal{H}}$ is defined by
\begin{equation}
\begin{aligned}
 \Tr_{\mathcal{H}} O=\sum_{i=0}^\infty
\bra i|O|i\ket.
\end{aligned} 
\end{equation}
The trace on the right hand side of \eqref{eq:TrA-Hilb}
can be simplified by 
using the following trick. We first notice that
\begin{equation}
\begin{aligned}
 ze^{z(a+a^\dagger)}=[a,e^{z(a+a^\dagger)}]=[e^{z(a+a^\dagger)},a^\dagger].
\end{aligned} 
\end{equation}
Then, using the relation
\begin{equation}
\begin{aligned}
 {[P,a]}=\rt{N}|N-1\ket \bra N|,\quad
{[a^\dagger,P]}=\rt{N}|N\ket \bra N-1|,
\end{aligned} 
\label{eq:Pcom}
\end{equation}
and the cyclicity of trace, we find
\begin{equation}
\begin{aligned}
 z\Tr A(z)&=\Tr_\mathcal{H} [a,e^{z(a+a^\dagger)}]P=\Tr_\mathcal{H} e^{z(a+a^\dagger)}[P,a]\\
&=\Tr_\mathcal{H} e^{z(a+a^\dagger)} \rt{N}|N-1\ket\bra N|
=\rt{N}\bra N|e^{z(a+a^\dagger)} |N-1\ket.
\end{aligned} 
\end{equation}
From the explicit form of matrix element in \eqref{eq:Aij} we 
arrive at the closed form of $\Tr A(z)$
\begin{equation}
\begin{aligned}
 \Tr A(z)=e^{\frac{z^2}{2}}L_{N-1}^1(-z^2).
\end{aligned} 
\end{equation}

Next consider the trace of the product of two $A(z)$'s
\begin{equation}
\begin{aligned}
 \Tr A(z_1)A(z_2)=\sum_{i,j=0}^{N-1}A(z_1)_{i,j}A(z_2)_{j,i}
=\Tr_\mathcal{H}\Bigl(e^{z_1(a+a^\dagger)}Pe^{z_2(a+a^\dagger)}P\Bigr).
\end{aligned} 
\label{eq:double}
\end{equation}
One can simplify this trace using the above trick 
by multiplying $z_1+z_2$
\begin{equation}
\begin{aligned}
 &(z_1+z_2)\Tr A(z_1)A(z_2)\\
=&\Tr_\mathcal{H}\Bigl( [a,e^{z_1(a+a^\dagger)}]P e^{z_2(a+a^\dagger)}P
+e^{z_1(a+a^\dagger)}P[a,e^{z_2(a+a^\dagger)}]P \Bigr)\\
=&\Tr_\mathcal{H}\Bigl(e^{z_1(a+a^\dagger)} [P,a] e^{z_2(a+a^\dagger)}P
+e^{z_1(a+a^\dagger)}P  e^{z_2(a+a^\dagger)} [P,a] \Bigr)\\
=&\rt{N}\bra N|\Bigl(e^{z_1(a+a^\dagger)} P e^{z_2(a+a^\dagger)}
+e^{z_2(a+a^\dagger)}Pe^{z_1(a+a^\dagger)}\Bigr)|N-1\ket.
\end{aligned} 
\label{eq:zmul}
\end{equation}
The last expression can be written as a single sum of matrix elements instead of the original
double sum \eqref{eq:double}
\begin{equation}
\begin{aligned}
 &(z_1+z_2)\Tr A(z_1)A(z_2)\\
=&\rt{N}\sum_{i=0}^{N-1}\Bigl[
A(z_1)_{i,N}A(z_2)_{i,N-1}+A(z_2)_{i,N}A(z_1)_{i,N-1}\Bigr]\\
=&e^{\frac{z_1^2+z_2^2}{2}}\sum_{i=0}^{N-1}\frac{i!}{(N-1)!}
\Bigl[z_1^{N-i}z_2^{N-1-i}L_i^{N-i}(-z_1^2)
L_{i}^{N-1-i}(-z_2^2)\\
&\hskip31mm +z_2^{N-i}z_1^{N-1-i}L_i^{N-i}(-z_2^2)
L_{i}^{N-1-i}(-z_1^2)\Bigr].
\end{aligned} 
\end{equation}
As far as we know, there is no formula to perform this summation
in a closed form.
However, it turns out that the derivative of this
expression can be written in a closed form. 

Let us act the derivative $\del_1-\del_2$ on the last expression in \eqref{eq:zmul} with
$\del_k=\frac{\del}{\del z_k}~(k=1,2)$. One can easily show that
\begin{equation}
\begin{aligned}
 &(z_1+z_2)(\del_1-\del_2)\Tr A(z_1)A(z_2)\\
=&\rt{N}\bra N|\Bigl(e^{z_1(a+a^\dagger)}[ a+a^\dagger,P] e^{z_2(a+a^\dagger)}
+e^{z_2(a+a^\dagger)}[P,a+a^\dagger]e^{z_1(a+a^\dagger)}\Bigr)|N-1\ket.
\end{aligned} 
\end{equation}
Again, using the relation \eqref{eq:Pcom}
this is simplified as
\begin{equation}
\begin{aligned}
&(z_1+z_2)(\del_1-\del_2)\Tr A(z_1)A(z_2)\\
=&N\Bigl[
\bra N|e^{z_1(a+a^\dagger)}|N\ket 
\bra N-1|e^{z_2(a+a^\dagger)}|N-1\ket -
\bra N-1|e^{z_1(a+a^\dagger)}|N-1\ket 
\bra N|e^{z_2(a+a^\dagger)}|N\ket
\Bigr]\\
=&Ne^{\frac{z_1^2+z_2^2}{2}}\Bigl[L_{N}(-z_1^2)L_{N-1}(-z_2^2)
-L_{N-1}(-z_1^2)L_{N}(-z_2^2)\Bigr] .
\end{aligned} 
\label{eq:TrAA-final}
\end{equation}
By setting $z_1=z$ and $z_2=\b{z}$ with $z,\b{z}$ defined in  
\eqref{eq:z-bt}, one can show that the above result \eqref{eq:TrAA-final}
leads to the exact form of $\del_t g_{\text{conn}}(\bt,t)$ in \eqref{eq:slope-bt}.


\end{document}